\documentclass[twocolumn,pra,showpacs,superscriptaddress,amssymb,amsmath,amsmath,floatfix]{revtex4-1}
\usepackage[T1]{fontenc}
\usepackage{graphicx}
\usepackage{epstopdf}
\usepackage{bm}
\usepackage[colorlinks=true,linkcolor=blue,urlcolor=blue,citecolor=blue]{hyperref}
\usepackage{comment}
\usepackage{cancel}
\usepackage{times}
\usepackage{amsmath, amsthm, amssymb, amsfonts}
\usepackage{xcolor}

\newcommand{\affFUW}{University of Warsaw, Faculty of Physics, ul. Pasteura 5, 02-093 Warsaw, Poland}
\newcommand{\affMIMUW}{University of Warsaw, Faculty of Mathematics, Informatics and Mechanics, ul. Banacha 2, 02-097 Warsaw, Poland}

\begin{document}
\author{Krzysztof My\'sliwy}
\affiliation{\affMIMUW}
\author{Piotr Wysocki}
\affiliation{\affFUW}
\author{Krzysztof Jachymski} 
\affiliation{\affFUW}
\email{krzysztof.jachymski@fuw.edu.pl}
\date{\today}
\title{No equivalence between hydrodynamic and dispersive mass of the charged polaron}
 \begin{abstract}
We consider the problem of a charged impurity exerting a weak, slowly decaying force on its surroundings, treating the latter as an ideal compressible fluid. In the semiclassical approximation, the ion is described by the Newton equation coupled to the Euler equation for the medium. After linearization, we obtain a simple closed formula for the effective mass of the impurity, depending on the interaction potential, the mean medium density, and sound velocity. Thus, once the interaction and the equation of state of the fluid is known, an estimate of the hydrodynamic effective mass can be quickly provided. Going beyond the classical case, we show that replacing the Newton with Schr\"{o}dinger equation can drastically change the behavior of the impurity. In particular, the scaling of the Fermi polaron effective mass with the medium density is opposite in quantum and classical scenario. Our results are relevant for experimental systems featuring low energy impurities in Fermi or Bose systems, such as ions immersed in neutral atomic gases. 
\end{abstract}
\maketitle

\section{Introduction}
Accelerating solid objects immersed in fluids experience additional inertia due to the induced motion of the surrounding fluid. The resulting resistance force, in contrast to the viscous drag, is at first order proportional to the body's acceleration. The proportionality coefficient might then be incorporated into the object's mass term in the relevant force balance, resulting in an \emph{effective mass}~\cite{landau1948effective,landau1987fluid} (also the terms \emph{added} or \emph{virtual mass} are used in the literature). The effective mass might significantly increase the body's inertia, which has important engineering applications e.g. in naval construction~\cite{Yeung1981,korotkin2008added}. The presence of this effect was realized quite early in the history of science, and it may be viewed as one of the first instances of renormalization. Analogous phenomena may also be found in the microscopic realm, with the effective mass of conduction electrons in solid state physics providing a prime well--known and important example~\cite{Kohn1957,kittel2018introduction}. The ubiquity of this effect across energy and length scales indicates that the added inertia due to interactions with the environment is expected to play a role in many fundamentally important systems such as nuclear matter~\cite{Jeukenne1976}, ion motion in liquid helium~\cite{Atkins1959,Reif1961}, transport of impurities in plasma~\cite{Hulse1983,Brown2005}, neutron stars~\cite{Pecak2024}.  It may also arise when manipulating microscopic objects for the purposes of nanotechnological or quantum engineering applications, for instance using ultracold gases~\cite{Massignan2014} or excitons in semiconductors~\cite{sidler2017fermi}. Modern quantum devices offer the opportunity to tune independently the properties of the medium and the interaction strength, representing an ideal testbed for benchmarking theoretical models as well as looking for new effects~\cite{Kohl2009,Catani2012,Chien2015,Dieterle2021}.

Various attempts, based on first--principle quantum mechanical calculations, have been employed to extract the value of the effective mass in different situations~\cite{Gross1962,rosch1999quantum}. The problem is inherently challenging, as it involves a nonequilibrium scenario and the structure of the medium can be nontrivial. Here, we would like to propose a very direct approach to this problem, in which the dominant effect affecting the impurity is simply the creation of sound waves in the medium. We expect this framework to work well in the case where the moving particle is relatively heavy, and its interaction with the medium is of \emph{long--range character} which cannot be modelled as a simple delta interaction. This is the case, for instance, for heavy ions moving in an environment of neutral atoms, with the atom-ion potential decaying with the fourth power of the distance~\cite{Tomza-2019}. In the ultracold regime, one may then expect the motion of the ion to be sufficiently slow to become dressed with the sound waves, with the ground state being of polaronic character~\cite{Astrakharchik2021,Christensen-2022,Mysliwy2024,Luis2024}. 

The main goal of this work is to analyze a comparatively simple system for which rigorous results may be derived, and establish a framework for further studies of more complex scenarios. to this end, we employ the hydrodynamic description of the medium. Its properties can be controlled by the form of the equation of state relating the pressure to the density.  The treatment of the ion is then twofold:
in the first part, we describe it as a classical particle described by Newton's equation, which is coupled to the underlying hydrodynamic equations governing the motion of the fluid which we linearize. We then perform an asymptotic analysis of the particle's motion, in the regime of small velocity as well as its derivatives, and for times $t$ large compared to the time it takes for a sound wave to pass the distance of the order of the range of the potential $t\gg L/c$.  As a result, we obtain the total force from the medium acting back on the particle which is proportional to the particle's acceleration. The added mass is then equal to the proportionality coefficient.  

In the second part, with the ultra--low temperature regime in mind, we change the description of the ion by assuming its motion is governed by a suitable Schrödinger equation, which is then coupled to the same set of hydrodynamic equations for the fluid. Analogous steps as those that lead us to the effective force in the classical case now result in an effective dynamics of the particle in the form a non--local and non--linear differential equation. Our approach to the effective mass problem in this regime is based on the following observation. Recall that the wave function of a free quantum mechanical particle is subject to dispersion in time, i.e., the variance of the position grows quadratically in time, with a coefficient inversely proportional to the particle's mass. Thus, we study the  dynamics numerically and look at the variance of the wave function in time. We verify that it is quadratic for weak potentials, at least for small times. This observation enables us to extract the \emph{dispersive effective mass} from the proportionality coefficient. Crucially, we find that it differs from the classical effective mass found in the previous treatment, as, in particular, it scales differently with the sound velocity. 




\section{Classical regime}
\subsection{The model and its linearization}
We consider a particle of mass $M$ moving in space filled with a fluid composed of particles of mass $m$ having local number  density $\rho(r,t)$ and pressure $P(\rho(r,t),T)$, where $T$ is the temperature, assumed constant throughout, and $r,t$ denote the spatial and time variable, respectively. We neglect the fluid viscosity and other dissipation effects. Far away from the impurity, whose position is labeled by $\vec{R}(t)$, the fluid density is homogeneous and equals $\rho_0>0$. 
The (isothermal) sound velocity in the fluid is denoted by 
\begin{equation}
c^2=\left.\left(\frac{\partial P}{\partial \rho}\right)_T\right|_{\rho=\rho_0}.
\end{equation}
The impurity interacts with the atoms of the fluid via a potential $V$, which we assume to vary slowly compared to the interatomic distance, but at the same time it decays reasonably fast at infinity so that both $V$ and its gradient are square integrable. The latter condition enables us to associate a length scale with the potential given by
\begin{equation}\label{L}
L=\frac{\int |\nabla V|^2 dr}{m^2c^4}.
\end{equation}

With $\vec{u}=\vec{u}(r,t)$ denoting the Euler velocity field of the fluid, we describe the system via coupling the continuity and Euler equations of the fluid to the Newton's equation for the particle as follows:
\begin{align}\label{en}
&\frac{\partial \vec{u}}{\partial t}+\vec{u}\cdot\nabla \vec{u}+\frac{1}{\rho}\left(\frac{\partial P}{\partial \rho}\right)_T\nabla \rho=-\frac{\nabla V(r-\vec{R}(t))}{m} \\
&\frac{\partial \rho}{\partial t}+\vec{u}\cdot\nabla \rho +\rho \nabla \cdot \vec{u}=0\\
&M\ddot{R} =\int \rho(r,t)\nabla V(r-R(t)) dr+f(R(t))\, , 
\end{align}
with $\nabla$ denoting the gradient with respect to $r$, the dot denoting time derivative, and $f$ stands for the external force acting on the particle. 
We are going to simplify the problem by assuming that 
\begin{itemize}
\item  the disturbances of $\rho$ due to the spatial homogeneity and time dependence are small compared to 
$\rho_0$, i.e. we write 
\begin{equation}
\rho=\rho_0+\phi(r,t), \quad |\phi|\ll \rho_0
\end{equation}
\item  the local fluid velocity is small compared to the speed of sound, $|u|\ll c$;
\item the particle's velocity is as well small compared to $c$, $\dot{R}\ll c$;
\item moreover, we assume that also the acceleration is small compared to the characteristic acceleration scale provided by $c$ and $L$ as defined by \eqref{L}, i.e., 
\begin{equation}
|\ddot{R}|\ll \frac{c^2}{L}
\end{equation}
and the acceleration fluctuations are small in the sense that also 
\begin{equation}
|\dddot{R}|\ll |\ddot{R}|\frac{c}{L}\ll \frac{c^3}{L^2}.
\end{equation}
\end{itemize}
We are interested in dynamics at times large enough such that $ct\gg L$ (note that this is consistent with the specified assumptions on the particle's velocity and its derivatives).
The assumptions on $\phi$ and $u$ enable us to perform standard linearization of the Euler equations. 
Retaining exclusively the terms that are linear in $\phi$ and $u$ in \eqref{en}, we obtain
\begin{align}
&\frac{\partial \vec{u}}{\partial t}+\frac{c^2}{\rho_0}\nabla \phi=-\frac{\nabla V(r-\vec{R}(t))}{m} \\
&\frac{\partial \phi}{\partial t}+\rho_0 \nabla \cdot \vec{u}=0\\
&M\ddot{R} =\int \phi(r,t)\nabla V(r-R(t)) dr+f(R(t)).
\end{align}
In the last line, we used that $V$ vanishes at infinity, and hence $\int \rho_0 \nabla V(r-R) dr=0$. 
Then we add the first equation to the second upon first acting with the time derivative on the second equation and the nabla on the first. This yields 
\begin{align}
&-\frac{1}{c^2} \frac{\partial^2 \phi}{\partial t^2}+\Delta \phi=-\frac{\rho_0}{mc^2}\Delta V(r-R)\\
&M\ddot{R} =\int \phi(r,t)\nabla V(r-R(t)) dr+f(R(t)).
\end{align}
As a check, let us first investigate the local disturbance of the fluid density close to the resting impurity and consider the static case $\phi(r,t)=\phi(r)$, $R=0, f=0$. From the first equation, it follows that $\phi+\frac{\rho_0}{mc^2} V(r)$ is a harmonic function. Since both $V$ and $\phi$ tend to zero far away from the impurity, $\phi+\frac{\rho_0}{mc^2} V(r)$ must be a constant equal to zero, and hence 
\begin{equation}\label{qtf}
\rho(r)=\rho_0\left(1-\frac{V(r)}{mc^2}\right).
\end{equation}
For concreteness, let us compare this with the case where the fluid is a zero--temperature ideal Fermi gas in $d$ dimensions. Thomas--Fermi theory yields then the static density profile~\cite{Mysliwy2024}
\begin{equation}\label{tf}
\rho(r)=\rho_0\left(\frac{\mu-V(r)}{\varepsilon_F}\right)^{\frac{d}{2}}
\end{equation}
with $\mu$ the chemical potential and $\varepsilon_F$ the Fermi level. At zero temperature, the sound velocity in the ideal Fermi gas is $c^2=\frac{2}{d}\frac{\varepsilon_F}{m}$. In the weakly interacting regime $\mu\approx \varepsilon_F$ and \eqref{tf} reduces to \eqref{qtf}, as expected. For consistency, we verify that the net force on the particle vanishes, $\int \rho(r) \nabla V(r) = \rho_0 \int \nabla V(r)-\frac{\rho_0}{mc^2}\int V(r)\nabla V(r) dr=-\frac{\rho_0}{2mc^2} \int \nabla V^2 dr=0$. 

Let us turn to the time--dependent case. We observe that $\eta(r,t)=\phi(r,t)+\frac{\rho_0}{mc^2}V(r-R(t))$ satisfies the inhomogeneous wave equation
\begin{equation}\label{inhW}
\frac{1}{c^2} \frac{\partial^2 \eta}{\partial t^2}-\Delta \eta=\frac{\rho_0}{mc^4}\frac{\partial^2 }{\partial t^2} V(r-R(t)).
\end{equation}
We take the static solution \eqref{qtf} as the initial condition, i.e. $\eta(r,0)=0$ and $\frac{\partial{\eta}}{\partial t}\eta(r,0)=0$. 
Then the solution of \eqref{inhW} is given by the \emph{Kirchhoff's formula}
\begin{equation}
\eta(r,t)=\frac{\rho_0}{4\pi mc^4}\int_{|r-r'|<ct} \frac{\partial^2 V}{\partial t^2} (r',ct-|r-r'|)\frac{1}{|r-r'|}dr'.
\end{equation}
Recalling that
\begin{equation}\label{V}
\frac{\partial^2 V}{\partial t^2}(r,t)=-\ddot{R}(t)\cdot \nabla V(r-R(t))+(\dot{R}(t)\cdot \nabla)^2 V(r-R(t))\, ,
\end{equation}
the total force from the medium hence equals 
\begin{align}
&\nonumber  F_m:=\int(\rho_0+ \phi(r,t))\nabla V(r-R(t)) dr= \\
&=\int \left( \eta(r,t)-\frac{\rho_0}{mc^2}V(r-R(t))\right) \nabla V(r-R(t)) dr.
\end{align}
Note that, by a computation identical to the static case, the total force from the \emph{adiabatic term} vanishes
\begin{align}
&\nonumber \int \frac{\rho_0}{mc^2}V(r-R(t)) \nabla V(r-R(t)) dr=\\ 
&=\int \frac{\rho_0}{2mc^2} \nabla V(r-R(t))^2 dr=0.
\end{align}
Plugging in the solution $\eta(r,t)$ into the remainder, we obtain
\begin{align}\label{force_raw}
F_m&=\frac{\rho_0}{4\pi mc^4}\int_{|r-r'|<ct}\frac{W_1(r,r',t)+W_2(r,r',t)}{|r-r'|}dr dr'
\end{align}
with 
\begin{equation}
W_1(r,r',t)=-\ddot{R}(t_c)\cdot \nabla V(r'-R(t_c))\nabla V(r-R(t))
\end{equation}
and
\begin{equation}
W_2(r,r',t)=(\dot{R}(t_c)\cdot \nabla)^2 V(r'-R(t_c))\nabla V(r-R(t)),
\end{equation}
where
\begin{equation}
t_c=t-\frac{|r-r'|}{c}.
\end{equation}
Now we make use of the assumption that the potential $V$, while long--ranged, does decay fast enough so that $V$ and its derivatives are negligibly small beyond the effective range $L$ defined in \eqref{L}. Accordingly, the dominant contribution in the integrals in \eqref{force_raw} arises from the regions $|r'|\leq L, |r|\leq  L$.  If we restrict attention to large enough times such that $ct \gg L$ we ensure that $|r-r'|\leq 2L \ll ct $ and, consequently, we can extend the integration limits to the entire space. Finally, we make use of the assumptions on the velocity and its derivatives: $\dot{R}\ll c$  $|\ddot{R}|\ll  c^2/L$ and $|\dddot{R}|\ll |\ddot{R}|c/L$  which enables us to replace $t_c \to t$ at leading order in both $\dot{R}(t_c)$ and $\ddot{R}(t_c)$. At the end of the day, we obtain a closed Newton's equation for the particle
\begin{align}
M\ddot{R}&=-\frac{\rho_0}{4\pi mc^4}\iint \frac{\ddot{R}\cdot \nabla^{'} V(r') \nabla V(r)}{|r-r'|}dr'dr\\ \nonumber 
&+\frac{\rho_0}{4\pi mc^4}\iint\frac{(\dot{R}\cdot \nabla^{'})^2 V(r') \nabla V(r)}{|r-r'|}dr dr'.
\end{align}
If we assume that the potential is spherically symmetric, then the second term vanishes (simply consider the change of variables $r,r'\to -r, -r'$, with $\nabla^{'}\to -\nabla^{'}, \nabla \to -\nabla$ ) and the first simplifies so that the equation can be rewritten as
\begin{equation}
M\ddot{R}=-\frac{\rho_0\ddot{R}}{4\pi mc^4}\iint \frac{\nabla^{'} V(r')\cdot \nabla V(r)}{|r-r'|}dr'dr. 
\end{equation}
This is proportional to the particle's acceleration, and hence yields the effective mass of the particle 
\begin{equation}\label{eq_mass}
M_{\rm{eff}}=M+\frac{\rho_0}{4\pi mc^4}\iint \frac{\nabla^{'} V(r')\cdot \nabla V(r)}{|r-r'|}dr'dr.
\end{equation}

\subsection{Discussion}
\subsubsection{Positivity}
Let us briefly discuss the resulting expression for the effective mass. We have $M_{\rm{eff}}>M$ by the well--known properties of the Coulomb self--interaction, $\iint g(r)g(r')|r-r'|^{-1}drdr'\geq 0$ for \emph{arbitrary} charge distributions $g$. Another way to see this is to use Fourier transforms, which yields the identity
\begin{equation}
\iint \frac{\nabla^{'} V(r')\cdot \nabla V(r)}{|r-r'|}dr'dr=4\pi \int |V(r)|^2 dr
\end{equation}
so that we can write \eqref{eq_mass} in equivalent form
\begin{equation}\label{eq_2}
M_{\rm{eff}}=M+\frac{\rho_0}{mc^4} \int |V(r)|^2 dr.
\end{equation}

\subsubsection{Incompressible limit: hard sphere}
Note that we have established \eqref{eq_2} under the assumptions $ct \gg L$ as well as $|u|\ll c$. In particular, this form is suitable for the study of the incompressible limit $c\to \infty$. A non--trivial result is in this case possible if we consider a sequence of potentials which depends on $c$ and becomes infinite as $c\to \infty$, in such a way that $\lim_{c\to \infty}\int |V|^2/c^4=$ is non--zero and finite. Consider, for instance 
\begin{equation}\label{Vc}
V(r)=\begin{cases}
\frac{1}{\sqrt{2}}mc^2, \quad  |r|\leq R \\
0, \quad |r|>R.
\end{cases}
\end{equation}
This model, in the $c\to\infty$ limit, yields a hard sphere potential of radius $R$, for which \eqref{eq_2} gives 
\begin{equation}
M_{\rm{eff}}=M+\frac{2\pi R^3}{3}m\rho_0=M+\frac{1}{2}M_{\rm{disp}}
\end{equation}
where $M_{\rm{disp}}$ denotes the mass of the fluid displaced by the sphere. This is a well--known result in hydrodynamics~\cite{landau1987fluid}, recovered here by the particular class of potentials \eqref{Vc}. 

\subsubsection{Low compressibility: degenerate Fermi gas}
Our second test of \eqref{eq_2} is the consistency of its predictions with known examples of fluids with high $c$. One such example is provided by the ideal Fermi gas at high density, for which $c^2=\frac{2}{3}\frac{\varepsilon_F}{m}\sim \rho^{2/3}$. The effective mass~\eqref{eq_2} thus scales with the density as $\rho/c^4\sim \rho^{-1/3}$. This may appear to contradict our intuition, as the effective mass becomes \emph{smaller} upon increasing the density of the Fermi gas, provided that the immersed particle behaves purely classically. This phenomenon of an effective decoupling of moving particles from dense Fermi gases is, however, known and its onset can be proven rigorously from first principles \cite{Jeblick2017}, where it is shown that for a sufficiently regular, not delta--like interaction, the impurity and the Fermi gas decouple for large $\rho$. In fact, it was observed already by Fermi and Teller that even if \emph{both} the impurity and the fermions are charged and, consequently, the underlying interaction is a Coulomb, the energy loss grows only logarithmically with the fermion density \cite{FermiTeller1947,  Mott1947,Arista1983}. This "low efficiency of slowing down" has important consequences e.g. in plasma physics \cite{Williams1975}. All these phenomena originate in the Pauli principle and can be understood in an elementary way by considering the energy and momentum transfer in individual collisions between the fermions and the impurity, as we argue below.  

Consider an ideal Fermi gas with one additional tracer particle. Due to the Pauli principle, only fermions with momenta close to the Fermi surface can participate in the collisions with the impurity, as the excitation of particle-hole pairs from deep below the surface would require large momenta. Since the typical energy transfer between the particles is of the order of the interaction $V$, fermions that 
occupy in momentum space the spherical shell of volume $ \Omega_V \sim p_F^2 dp =2 m V p_F$, where $p_F$ is the Fermi momentum, and $dp$ is the width of the shell, estimated from the energy transfer $(p_F+dp)^2-p_F^2\sim 2m V$. This yields the fraction of active fermions in momentum space
\begin{equation}
f=\frac{\Omega_V}{\Omega_F}\sim \frac{mV}{p_F^2}
\end{equation}
 with $\Omega_F$ being the volume of the Fermi ball. The number of encountered particles per unit volume $N_c$ is proportional to the number density of fermions $N\sim \rho_0$, with $f N_c$ among them leading to actual momentum transfer. The latter can be estimated as 
 \begin{equation}
 \delta p= F \tau_c
 \end{equation}
 with $F$ the force and $\tau_c$ the typical collision time, which is effectively very short due to the large momentum of the fermions involved and can be estimated as $\tau_c=\frac{mR}{|p_F+mu|}$, where $u$ is the velocity of the impurity and $R$ is the typical scale on which the potential varies, so that also $F\sim V/R$. The net momentum transfer averaged over directions can be roughly estimated by subtracting  heads-on and tails-on collisions, i.e., along the direction of $u$: 
\begin{equation}
\delta\bar{p}\approx F \left(\frac{mR}{p_F-mu}-\frac{mR}{p_F+mu}\right)\sim \frac{2m^2V}{p_F^2}u.
\end{equation}
Thus the total net average momentum transfer per volume is 
\begin{equation}
\delta P=f N_c \delta\bar{p}\approx \frac{\rho_0}{m} \frac{mV}{p_F^2}\frac{2m^2V}{p_F^2}u\sim \frac{\rho_0m^2 V^2}{p_F^4}u
\end{equation} 
which decays as $\rho_0^{-1/3}$ with the density in agreement with the effective mass formula found above. We note that the collision estimates are based on the classical behavior of the particles, which is justified for the colliding fermions due to their speed, but works for the impurity only by assumption.
Our next step is to investigate the consequences of lifting this assumption. We shall see that the behavior of the effective mass becomes substantially different in the quantum regime. This is due to the additional energy scale of the particle $\frac{\hbar^2}{2MR^2}$ present only in the quantum regime which affects the momentum transfer, leading to an \emph{increase} of the effective mass with the density.

\section{Quantum mechanical treatment of the impurity}
We now turn to the quantum mechanical description of the particle, i.e., instead of coupling the Euler equation of the fluid to Newton's equation of the particle, we replace the latter with a suitable Schrödinger equation. In particular, we replace the deterministic potential acting on the fluid $V(r-R(t)$, where $R(t)$ is the instanteneous position of the particle, with its quantum--mechanical analogue  $\int V(r-y) |\psi(y,t)|^2dy$, where $\psi(y,t)$ is the time--dependent wave function of the impurity, with its square yielding the probability density of the instantaneous position of the particle. The coupled Euler--Schrödinger equations read 
\begin{align}\label{en2}
&\frac{\partial \vec{u}}{\partial t}+\vec{u}\cdot\nabla \vec{u}+\frac{1}{\rho}\left(\frac{\partial P}{\partial \rho}\right)_T\nabla \rho=-\frac{\nabla }{m} \int V(r-y) |\psi(y,t)|^2dy \\
 &\frac{\partial \rho}{\partial t}+\vec{u}\cdot\nabla \rho +\rho \nabla \cdot \vec{u}=0\\
&i\hbar \frac{\partial \psi(x,t)}{\partial t}=-\frac{\hbar^2}{2M}\Delta \psi(x,t) +\int \rho(y,t)( V(x-y) \psi(x,t)dy
\end{align}
After performing the linearization of the fluid part of the system as in the previous section, we obtain the wave equation
\begin{equation}
-\frac{1}{c^2} \frac{\partial^2 \phi}{\partial t^2}+\Delta \phi=-\frac{\rho_0}{mc^2}\Delta V_{\psi_t}(r)
\end{equation}
where we adopted the notation 
\begin{equation}
V_{\psi_t}(r)=\int V(r-y) |\psi(y,t)|^2dy.
\end{equation}
We observe that the solutions look exactly the same as in the classical case under the replacement $V\rightarrow V_{\psi_t}$, and we can write
\begin{align}\label{ff}
\phi(r,t)&=-\frac{\rho_0}{mc^2}V_{\psi_t}+\\ \nonumber+
&\frac{\rho_0}{4\pi mc^4}\int_{|r-r'|<ct} \frac{\partial^2 V_{\psi_t}}{\partial t^2} (r',ct-|r-r'|)\frac{1}{|r-r'|}dr'.
\end{align}
In particular, we immediately recover the classical case by setting $|\psi(y,t)|^2=\delta(y-R(t))$ as expected.

In the classical case, we have seen that the \emph{adiabatic} term $-\frac{\rho_0}{mc^2}V(r-R(t))$, corresponding to the first term in Eq. \eqref{ff} above, does not produce any net force acting on the particle, and the effective mass scales as $\sim c^{-4}$ with the sound velocity at leading order. In the Fermi gas case, this resulted in the scaling of the effective mass as $\rho_0^{-1/3}$, in agreement with the kinematic considerations and the phenomenon of an effective decoupling of slow impurities from dense Fermi gases as observed in the motion of ions in plasma. 

Now we shall show that the adiabatic term, i.e., the first term on the RHS of \eqref{ff} does not in general result in a vanishing force when quantum effects are not to be neglected in the motion of the particle. This leads to a different scaling of the effective mass with the density. According to Ehrenfest theorem, we can compute the mean force acting on the particle as 
\begin{equation}
\iint |\psi(y,t)|^2 m^{-1}\rho(r,t) \nabla_y V(r-y) dr dy
\end{equation}
We investigate the effect of the adiabatic term $-\frac{\rho_0}{mc^2}V_{\psi_t}$. After simple manipulations using $ \nabla_y V(r-y)=-\nabla_r V(r-y),$ the resulting force can be written as
\begin{equation}
\frac{\rho_0}{m^2c^2}\iiint |\psi(y,t)|^2 |\psi(z+y,t)|^2 V(r-z) \nabla_r V(r) dr dy dz.
\end{equation}
Again, for $|\psi(y,t)|^2=\delta(y-R(t))$, the force vanishes due to $\int V(r)\nabla_r V(r) dr=0$. However, for a general position probability density, the latter integral becomes 
\begin{equation}
\int V(r-z)\nabla_r V(r) dr=\nabla_z \int |\hat{V}(k)|^2 e^{ikz } dk \equiv F(z)
\end{equation} 
where we used the Parseval's identity $\int \overline{f}g =\int \overline{\hat{f}}\hat {g}$, with the hat denoting the Fourier transform. Note that $F(z)$ is in general non--zero, while $F(0)$, which is the only value appearing in the classical case, is zero because $\hat f(0)\geq \hat{f}(k)$ for any Fourier transform of a positive function, and thus also $(\nabla \hat{f})(0)=0$. Thus, finally, the net average force on a particle in state $\psi(y,t)$ stemming from the adiabatic term has the form
\begin{equation}\label{fad}
\frac{\rho_0}{m^2c^2}\iint |\psi(y,t)|^2 |\psi(z,t)|^2 F(y-z) dy dz.
\end{equation}
which scales as $\rho_0/c^2$ and is in general non--vanishing, in contrast to the classical case. We emphasize that the origin of this force is rooted in the fluctuations of the positions as it vanishes for a delta distribution. In the next step, we shall have a closer look on the effect of the \emph{adiabatic} force \eqref{fad} on the effective mass. 

\subsection{Dispersive effective mass}
\subsubsection{Free evolution}
We investigate the Schrödinger equation arising from evaluating the potential resulting from the adiabatic part of the density profile $-\frac{\rho_0}{m^2c^2}V_{\psi_t}$, which has the form
\begin{align}\label{scr}
&i\hbar \frac{\partial \psi(x,t)}{\partial t}=-\frac{\hbar^2}{2M}\Delta \psi(x,t)+ \\ \nonumber 
& -\frac{\rho_0}{m^2c^2}\left(\iint V(x-z)V(z-y)|\psi(y,t)|^2\right)dydz \psi(x,t).
\end{align}
This nonlocal nonlinear equation has a similar form to the well-known Gross-Pitaevskii equation. Our approach to understand its properties will be based on looking at the dynamical spreading of wave packets.
We first recall this behavior for free wave packets, i.e. for $V=0$. Assume that the initial wave function has a momentum distribution satisfying $\langle xp \rangle(0) =0$ (no squeezing)).
The evolution of $\langle x^2\rangle(t)=\langle  \psi_t |x^2\psi_t\rangle$ is governed by the Heisenberg equation
\begin{equation}
	\frac{d}{dt}\langle x^2\rangle(t)=\frac{1}{i\hbar}\langle [x^2,H]\rangle =\frac{1}{2i\hbar m} \langle [x^2,p^2] \rangle.
\end{equation}
 We have $[x^2,p^2]=x([x,p]p+p[x,p])x$ and hence
 \begin{equation}
 \frac{d}{dt}\langle x^2\rangle(t)=
 \frac{1}{2i\hbar m} \langle [x^2,p^2]=\frac{1}{2m}\langle xp+px \rangle = \frac{2}{m}\mathrm{Re} \langle xp \rangle.
 \end{equation}
 On the other hand, 
 \begin{equation}
 	\frac{d}{dt}\langle xp \rangle=\frac{1}{2i\hbar m}\frac{d}{dt} \langle [xp,p^2]\rangle=\frac{\langle p^2\rangle }{m}. 
 \end{equation}
 The last term is real and constant in time since $[H,p^2]=0$ by assumption. Thus 
 \begin{equation}
 	\langle p x \rangle = \frac{\langle p^2\rangle }{m}t+\langle p x\rangle(0).
 \end{equation}
 The last term vanishes due to our non--squeezing assumption, and finally 
 \begin{equation}
 	\frac{d}{dt}\langle x^2\rangle(t)=\frac{2}{m^2}\langle p^2 \rangle(0)t
 \end{equation} 
We see that regardless of the initial form of the wave packet, provided that the no-squeezing condition is satisfied, the variance spreads quadratically in time, with proportionality coefficient scaling with the inverse square mass of the particle. It is easy to check that if one does not assume $\langle x p\rangle =0$ at the beginning, then the same conclusion follows for times long enough, in the sense that the term $\langle x p\rangle $ rises only linearly in time and is negligible for larger times. This simple result is the basis for the estimation of the effective mass in the interacting case, as discussed in the next paragraph.

\subsubsection{Evolution with the self--interacting term}
We now wish to investigate the hypothesis that after including the adiabatic non--local interaction depicted in, it is still true that 
\begin{equation}\label{eq.spreading}
	\langle x^2 \rangle (t) \approx \frac{\langle p^2 \rangle (0)}{(M^*)^2}t^2 + \langle x^2 \rangle(0)\, .
\end{equation}
At this point we will turn to numerical solution of the nonlinear differential equation~\eqref{scr}. To this end, we fix the form of the potential to 
\begin{equation}
V(x)=-\frac{C_4}{(b^2+x^2)^2}
\end{equation}
which is a common choice for modelling the interaction between ions and neutral atoms~\cite{Tomza-2019}. 
The regularizing parameter $b>0$ can be thought as being subject to experimental control via a Feshbach resonance and controls the scattering length and the bound state energy of the potential.
With this choice, we set the unit of length $\lambda=\sqrt{\frac{C_4M}{\hbar^2}}$ and time $\tau=\frac{M\lambda^2}{\hbar}$, so that equation~\eqref{scr} assumes the dimensionless form
\begin{align}
&i\frac{\partial}{\partial t} \psi(x,t)=-\frac{1}{2}\Delta \psi(x,t)\\  \nonumber 
&-g\left(\iint  V^{(2)}(x,y,z;b)|\psi(y,t)|^2 dydz\right) \psi(x,t)
\end{align}
with
\begin{equation}
V^{(2)}(x,y,z;b)=\frac{1}{(b^2+(x-z)^2)^2}\frac{1}{(b^2+(z-y)^2)^2}
\end{equation}
and the dimensionless coupling constant 
\begin{equation}\label{eq:g}
g=\frac{\rho_0}{mc^2}\sqrt{\frac{C_4\hbar^2}{M}}.
\end{equation}
We now set the value of $b$ to $b=1.2\lambda$, such that the potential is indeed weak and slowly varying. With such units, Eq.~\eqref{eq.spreading} becomes
\begin{equation}\label{eq.spreading.dimless}
    \langle x^2\rangle(t) \approx \left(\frac{M}{M^*}\right)^2\langle p^2\rangle(0)t^2 + \langle x^2\rangle(0),
\end{equation}
with the rate of spreading encoded in the ratio $M/M^*$.

\begin{figure}
	\includegraphics[width=0.95\linewidth]{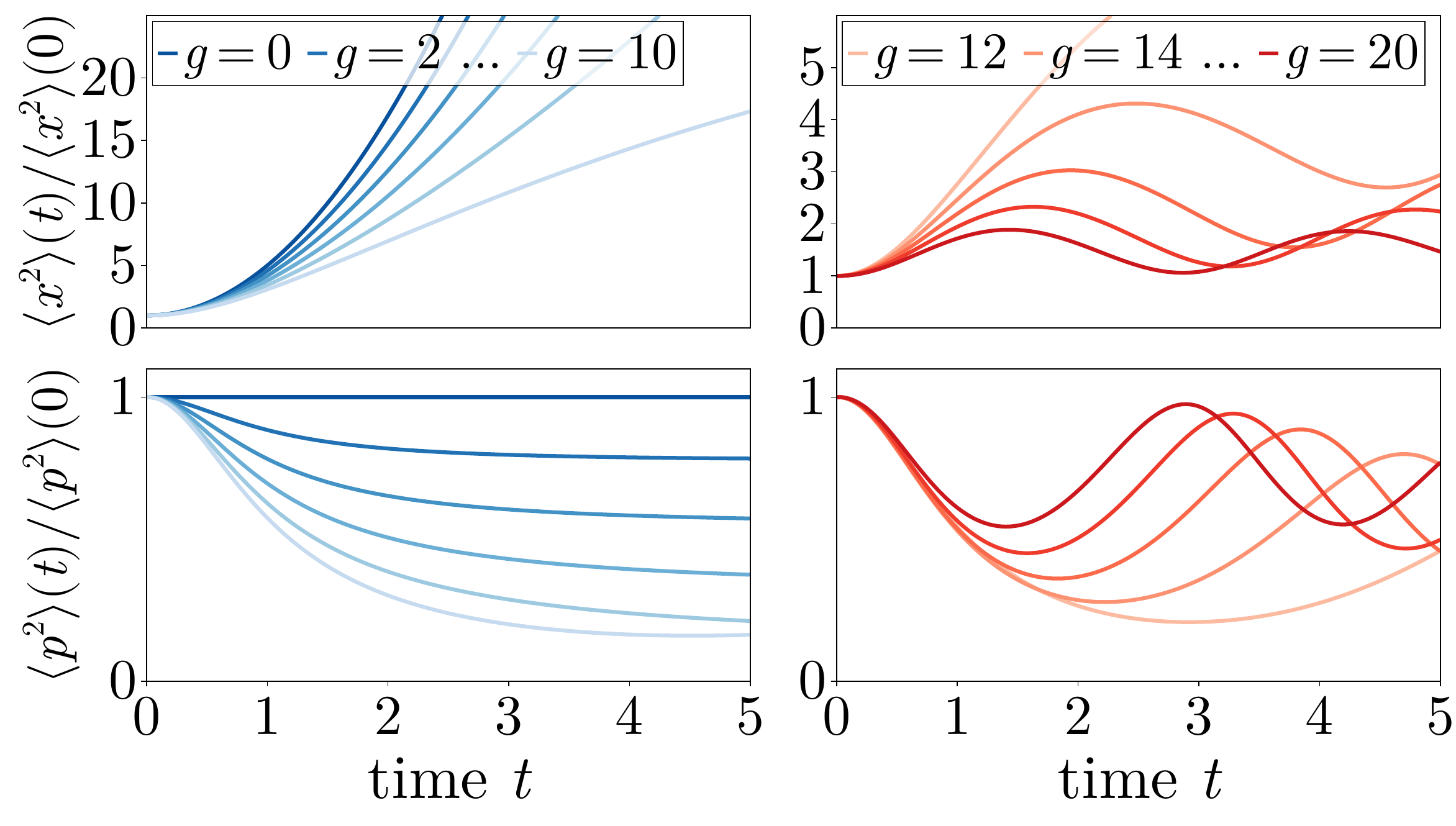}
	\caption{\label{var_and_p} Upper row: variance of the position of the impurity for weak (left) and strong (right) interactions controlled by the coupling constant $g$ defined in eq.~\eqref{eq:g}. Note the change of vertical scale between the panels. Lower row: same, but for the impurity momentum.}
\end{figure}

\begin{figure}[t]
	\includegraphics[width=0.95\linewidth]{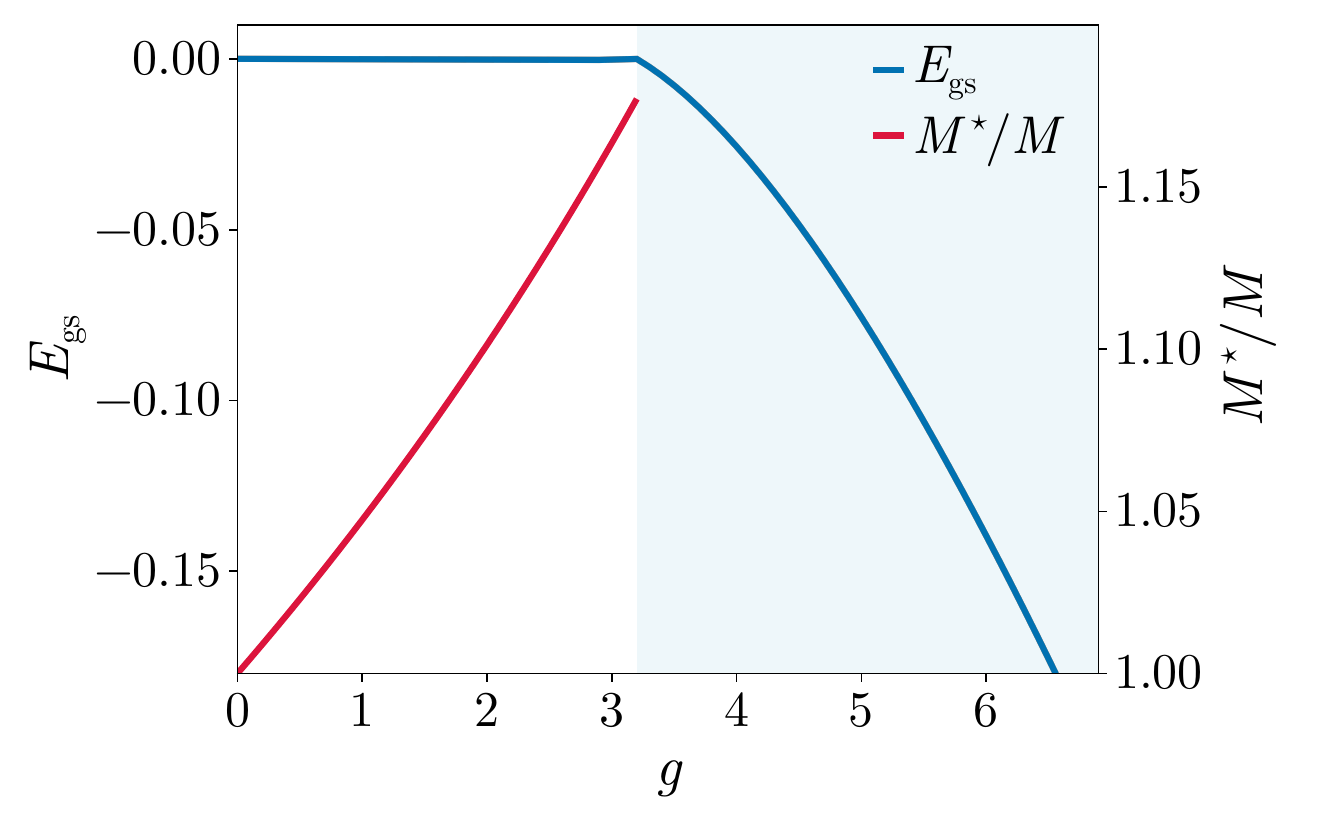}
	\caption{\label{en_v_g} Energy (blue) and effective mass (red) of the impurity as a function of the interaction strength $g$. Shaded area corresponds to a self-bound state.}
\end{figure}

With the form of $V$ fixed, we numerically find the evolved wave function from~\eqref{scr}, and  compute its variance $\langle x^2 \rangle(t)$ as a function of time, for different values of the coupling $g$. The results are presented in Fig. \ref{var_and_p}. We observe that for small enough $g$, the growth of the variance is approximately parabolic in time even for large times. In contrast, for larger $g$, the parabolic shape is quickly replaced by oscillatory behavior, which marks the departure from the quasi--free dispersive evolution of the wave function. This crossover is to be expected in view of the occurrence of the localization transition in the static model, i.e., the appearance of a bound state for larger values of~$g$ \cite{Mysliwy2024}. For strong interactions the impurity is no longer spreading, but rather oscillates in a way that reminds of a breathing mode of a trapped gas~\cite{Grochowski2020}. Similar behavior is observed for the variance of the momentum. For no interactions the wavepacket retains its shape such that it stays constant, while the nonlinear term tends to localize the particle in momentum space, reflecting the position spreading. Again, for strong interactions we enter a qualitatively different regime where oscillatory dynamics is observed. This transition is very well reflected by the behaviour of the energy depicted in Fig. \ref{en_v_g}. For weak interaction the ground state energy is zero, but already for $g\gtrsim 3$ a transition to a bound state takes place. As the particle is no longer free, it does not make sense to speak about its effective mass in this regime.

\begin{figure}[b]
	\includegraphics[width=0.95\linewidth]{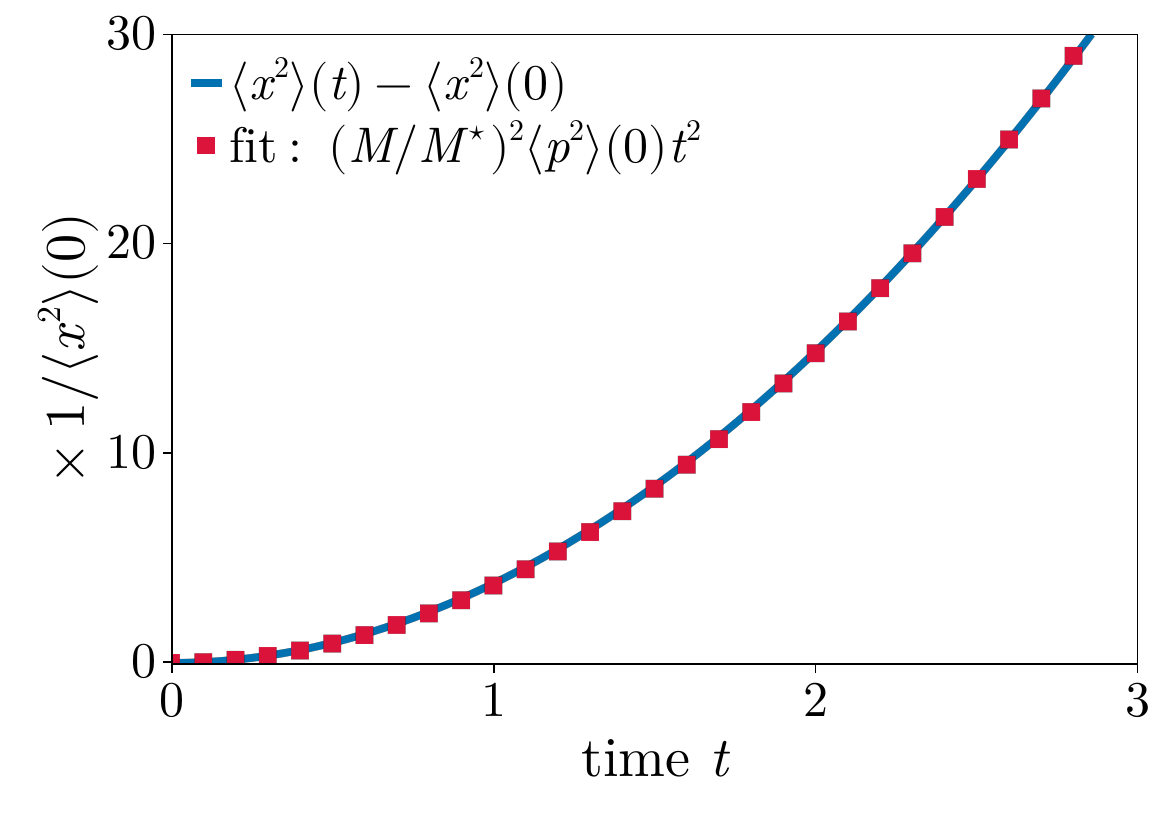}
	\caption{\label{en_fit} Example of fitting of the formula~\eqref{eq.spreading.dimless} to the numerical results, which for small enough $g$ turns out to be very precise.}
\end{figure}

In the next step, we choose values of $g$ such that the variance is quadratic in time for sufficiently large times, such that $\langle x^2 \rangle (t)= A\left(t\right)^2+o\left(\frac{t}{T}\right)^2$.
We observe that in this regime, the variance of the momentum is changing on a much slower scale, cf. Fig. \ref{var_and_p}.
Thus, comparison with eq.~\eqref{eq.spreading.dimless} allows us to define the dimensionless dispersive effective mass as follows
\begin{equation}
    \frac{M^*}{M} = \sqrt{\frac{\langle p^2\rangle(0)}{A}}
\end{equation}
for the given initial momentum distribution of the wave packet $\langle p^2(0)\rangle$. The value of the so--defined effective mass may then be obtained from the value of $A$, extracted by a polynomial fit  after subtracting the initial position variance $\langle x^2\rangle(0)$. In Fig. \ref{en_v_g} we present the obtained values of the dimensionless dispersive mass $M^*/M$ against $g$. As shown in Fig.~\ref{en_fit}, the fit quality is excellent. In particular, we verify that $M^*$ is linear in $g$ for small $g$, i.e., 
\begin{equation}
\frac{M^*-M}{M}\approx C_b g=C_b\frac{\rho_0}{mc^2}\sqrt{\frac{C_4\hbar^2}{M}}\, 
\end{equation}
with a constant $C_b>0$ depending on $b$ (e.g. for the case $b=1.2\lambda$ considered here, $C_b=0.045$). 
In particular, with other parameters fixed, $M^*$ decays with the sound velocity as $c^{-2}$ and not as $c^{-4}$, as was the case for the classical effective mass found in eq.~\eqref{eq_2}. The dependence on $\hbar$ marks here the additional energy scale set by the position fluctuations. For the case of the ideal Fermi gas where $c^2\sim \rho^{2/3}$ we find that now $M^*\sim \rho^{1/3}$ so that the dispersive mass \emph{grows} with the density, in contrast to the dynamical effective mass which decayed with the density as $\rho^{-1/3}$. 

\section{Conclusions}
We have presented a direct approach to estimating the effective mass of the impurity in an ideal fluid. For an ideal Fermi gas, we were able to analytically predict the scaling with the gas density in the perturbative limit, and identify the source of the discrepancy between the classical and quantum prediction, which is rooted in finite width of the position distribution given by the impurity wave function. This leads not only to the lack of equivalence $M^*\neq M_{\rm{eff}}$, but to strikingly different scaling of the quantum and classical added mass with the fluid density.

\emph{Acknowledgements.} This work was supported by the National Science Centre of Poland Grant 2020/37/B/ST2/00486. K.M. would like to thank P. B. Mucha for fruitful suggestions regarding the asymptotic analysis of the classical case. 

\bibliography{refs}

\begin{thebibliography}{31}%
\makeatletter
\providecommand \@ifxundefined [1]{%
 \@ifx{#1\undefined}
}%
\providecommand \@ifnum [1]{%
 \ifnum #1\expandafter \@firstoftwo
 \else \expandafter \@secondoftwo
 \fi
}%
\providecommand \@ifx [1]{%
 \ifx #1\expandafter \@firstoftwo
 \else \expandafter \@secondoftwo
 \fi
}%
\providecommand \natexlab [1]{#1}%
\providecommand \enquote  [1]{``#1''}%
\providecommand \bibnamefont  [1]{#1}%
\providecommand \bibfnamefont [1]{#1}%
\providecommand \citenamefont [1]{#1}%
\providecommand \href@noop [0]{\@secondoftwo}%
\providecommand \href [0]{\begingroup \@sanitize@url \@href}%
\providecommand \@href[1]{\@@startlink{#1}\@@href}%
\providecommand \@@href[1]{\endgroup#1\@@endlink}%
\providecommand \@sanitize@url [0]{\catcode `\\12\catcode `\$12\catcode
  `\&12\catcode `\#12\catcode `\^12\catcode `\_12\catcode `\%12\relax}%
\providecommand \@@startlink[1]{}%
\providecommand \@@endlink[0]{}%
\providecommand \url  [0]{\begingroup\@sanitize@url \@url }%
\providecommand \@url [1]{\endgroup\@href {#1}{\urlprefix }}%
\providecommand \urlprefix  [0]{URL }%
\providecommand \Eprint [0]{\href }%
\providecommand \doibase [0]{http://dx.doi.org/}%
\providecommand \selectlanguage [0]{\@gobble}%
\providecommand \bibinfo  [0]{\@secondoftwo}%
\providecommand \bibfield  [0]{\@secondoftwo}%
\providecommand \translation [1]{[#1]}%
\providecommand \BibitemOpen [0]{}%
\providecommand \bibitemStop [0]{}%
\providecommand \bibitemNoStop [0]{.\EOS\space}%
\providecommand \EOS [0]{\spacefactor3000\relax}%
\providecommand \BibitemShut  [1]{\csname bibitem#1\endcsname}%
\let\auto@bib@innerbib\@empty
\bibitem [{\citenamefont {Landau}\ and\ \citenamefont
  {Pekar}(1948)}]{landau1948effective}%
  \BibitemOpen
  \bibfield  {author} {\bibinfo {author} {\bibfnamefont {L.}~\bibnamefont
  {Landau}}\ and\ \bibinfo {author} {\bibfnamefont {S.}~\bibnamefont {Pekar}},\
  }\href@noop {} {\bibfield  {journal} {\bibinfo  {journal} {Zh. Eksp. Teor.
  Fiz}\ }\textbf {\bibinfo {volume} {18}},\ \bibinfo {pages} {419} (\bibinfo
  {year} {1948})}\BibitemShut {NoStop}%
\bibitem [{\citenamefont {Landau}\ and\ \citenamefont
  {Lifshitz}(1987)}]{landau1987fluid}%
  \BibitemOpen
  \bibfield  {author} {\bibinfo {author} {\bibfnamefont {L.~D.}\ \bibnamefont
  {Landau}}\ and\ \bibinfo {author} {\bibfnamefont {E.~M.}\ \bibnamefont
  {Lifshitz}},\ }\href@noop {} {\emph {\bibinfo {title} {Fluid Mechanics:
  Volume 6}}},\ Vol.~\bibinfo {volume} {6}\ (\bibinfo  {publisher} {Elsevier},\
  \bibinfo {year} {1987})\BibitemShut {NoStop}%
\bibitem [{\citenamefont {Yeung}(1981)}]{Yeung1981}%
  \BibitemOpen
  \bibfield  {author} {\bibinfo {author} {\bibfnamefont {R.~W.}\ \bibnamefont
  {Yeung}},\ }\href@noop {} {\bibfield  {journal} {\bibinfo  {journal} {Applied
  Ocean Research}\ }\textbf {\bibinfo {volume} {3}},\ \bibinfo {pages} {119}
  (\bibinfo {year} {1981})}\BibitemShut {NoStop}%
\bibitem [{\citenamefont {Korotkin}(2008)}]{korotkin2008added}%
  \BibitemOpen
  \bibfield  {author} {\bibinfo {author} {\bibfnamefont {A.~I.}\ \bibnamefont
  {Korotkin}},\ }\href@noop {} {\emph {\bibinfo {title} {Added masses of ship
  structures}}},\ Vol.~\bibinfo {volume} {88}\ (\bibinfo  {publisher} {Springer
  Science \& Business Media},\ \bibinfo {year} {2008})\BibitemShut {NoStop}%
\bibitem [{\citenamefont {Kohn}(1957)}]{Kohn1957}%
  \BibitemOpen
  \bibfield  {author} {\bibinfo {author} {\bibfnamefont {W.}~\bibnamefont
  {Kohn}},\ }\href {\doibase 10.1103/PhysRev.105.509} {\bibfield  {journal}
  {\bibinfo  {journal} {Phys. Rev.}\ }\textbf {\bibinfo {volume} {105}},\
  \bibinfo {pages} {509} (\bibinfo {year} {1957})}\BibitemShut {NoStop}%
\bibitem [{\citenamefont {Kittel}\ and\ \citenamefont
  {McEuen}(2018)}]{kittel2018introduction}%
  \BibitemOpen
  \bibfield  {author} {\bibinfo {author} {\bibfnamefont {C.}~\bibnamefont
  {Kittel}}\ and\ \bibinfo {author} {\bibfnamefont {P.}~\bibnamefont
  {McEuen}},\ }\href@noop {} {\emph {\bibinfo {title} {Introduction to solid
  state physics}}}\ (\bibinfo  {publisher} {John Wiley \& Sons},\ \bibinfo
  {year} {2018})\BibitemShut {NoStop}%
\bibitem [{\citenamefont {Jeukenne}\ \emph {et~al.}(1976)\citenamefont
  {Jeukenne}, \citenamefont {Lejeune},\ and\ \citenamefont
  {Mahaux}}]{Jeukenne1976}%
  \BibitemOpen
  \bibfield  {author} {\bibinfo {author} {\bibfnamefont {J.~P.}\ \bibnamefont
  {Jeukenne}}, \bibinfo {author} {\bibfnamefont {A.}~\bibnamefont {Lejeune}}, \
  and\ \bibinfo {author} {\bibfnamefont {C.}~\bibnamefont {Mahaux}},\
  }\href@noop {} {\bibfield  {journal} {\bibinfo  {journal} {Physics Reports}\
  }\textbf {\bibinfo {volume} {25}},\ \bibinfo {pages} {83} (\bibinfo {year}
  {1976})}\BibitemShut {NoStop}%
\bibitem [{\citenamefont {Atkins}(1959)}]{Atkins1959}%
  \BibitemOpen
  \bibfield  {author} {\bibinfo {author} {\bibfnamefont {K.~R.}\ \bibnamefont
  {Atkins}},\ }\href {\doibase 10.1103/PhysRev.116.1339} {\bibfield  {journal}
  {\bibinfo  {journal} {Phys. Rev.}\ }\textbf {\bibinfo {volume} {116}},\
  \bibinfo {pages} {1339} (\bibinfo {year} {1959})}\BibitemShut {NoStop}%
\bibitem [{\citenamefont {Meyer}\ and\ \citenamefont {Reif}(1961)}]{Reif1961}%
  \BibitemOpen
  \bibfield  {author} {\bibinfo {author} {\bibfnamefont {L.}~\bibnamefont
  {Meyer}}\ and\ \bibinfo {author} {\bibfnamefont {F.}~\bibnamefont {Reif}},\
  }\href {\doibase 10.1103/PhysRev.123.727} {\bibfield  {journal} {\bibinfo
  {journal} {Phys. Rev.}\ }\textbf {\bibinfo {volume} {123}},\ \bibinfo {pages}
  {727} (\bibinfo {year} {1961})}\BibitemShut {NoStop}%
\bibitem [{\citenamefont {Hulse}(1983)}]{Hulse1983}%
  \BibitemOpen
  \bibfield  {author} {\bibinfo {author} {\bibfnamefont {R.~A.}\ \bibnamefont
  {Hulse}},\ }\href@noop {} {\bibfield  {journal} {\bibinfo  {journal} {Nuclear
  Technology-Fusion}\ }\textbf {\bibinfo {volume} {3}},\ \bibinfo {pages} {259}
  (\bibinfo {year} {1983})}\BibitemShut {NoStop}%
\bibitem [{\citenamefont {Brown}\ \emph {et~al.}(2005)\citenamefont {Brown},
  \citenamefont {Preston},\ and\ \citenamefont {Singleton~Jr}}]{Brown2005}%
  \BibitemOpen
  \bibfield  {author} {\bibinfo {author} {\bibfnamefont {L.~S.}\ \bibnamefont
  {Brown}}, \bibinfo {author} {\bibfnamefont {D.~L.}\ \bibnamefont {Preston}},
  \ and\ \bibinfo {author} {\bibfnamefont {R.~L.}\ \bibnamefont
  {Singleton~Jr}},\ }\href@noop {} {\bibfield  {journal} {\bibinfo  {journal}
  {Physics Reports}\ }\textbf {\bibinfo {volume} {410}},\ \bibinfo {pages}
  {237} (\bibinfo {year} {2005})}\BibitemShut {NoStop}%
\bibitem [{\citenamefont {P\ifmmode~\mbox{\k{e}}\else \k{e}\fi{}cak}\ \emph
  {et~al.}(2024)\citenamefont {P\ifmmode~\mbox{\k{e}}\else \k{e}\fi{}cak},
  \citenamefont {Zdanowicz}, \citenamefont {Chamel}, \citenamefont
  {Magierski},\ and\ \citenamefont {Wlaz\l{}owski}}]{Pecak2024}%
  \BibitemOpen
  \bibfield  {author} {\bibinfo {author} {\bibfnamefont {D.}~\bibnamefont
  {P\ifmmode~\mbox{\k{e}}\else \k{e}\fi{}cak}}, \bibinfo {author}
  {\bibfnamefont {A.}~\bibnamefont {Zdanowicz}}, \bibinfo {author}
  {\bibfnamefont {N.}~\bibnamefont {Chamel}}, \bibinfo {author} {\bibfnamefont
  {P.}~\bibnamefont {Magierski}}, \ and\ \bibinfo {author} {\bibfnamefont
  {G.}~\bibnamefont {Wlaz\l{}owski}},\ }\href {\doibase
  10.1103/PhysRevX.14.041054} {\bibfield  {journal} {\bibinfo  {journal} {Phys.
  Rev. X}\ }\textbf {\bibinfo {volume} {14}},\ \bibinfo {pages} {041054}
  (\bibinfo {year} {2024})}\BibitemShut {NoStop}%
\bibitem [{\citenamefont {Massignan}\ \emph {et~al.}(2014)\citenamefont
  {Massignan}, \citenamefont {Zaccanti},\ and\ \citenamefont
  {Bruun}}]{Massignan2014}%
  \BibitemOpen
  \bibfield  {author} {\bibinfo {author} {\bibfnamefont {P.}~\bibnamefont
  {Massignan}}, \bibinfo {author} {\bibfnamefont {M.}~\bibnamefont {Zaccanti}},
  \ and\ \bibinfo {author} {\bibfnamefont {G.~M.}\ \bibnamefont {Bruun}},\
  }\href@noop {} {\bibfield  {journal} {\bibinfo  {journal} {Rep. Prog. Phys.}\
  }\textbf {\bibinfo {volume} {77}},\ \bibinfo {pages} {034401} (\bibinfo
  {year} {2014})}\BibitemShut {NoStop}%
\bibitem [{\citenamefont {Sidler}\ \emph {et~al.}(2017)\citenamefont {Sidler},
  \citenamefont {Back}, \citenamefont {Cotlet}, \citenamefont {Srivastava},
  \citenamefont {Fink}, \citenamefont {Kroner}, \citenamefont {Demler},\ and\
  \citenamefont {Imamoglu}}]{sidler2017fermi}%
  \BibitemOpen
  \bibfield  {author} {\bibinfo {author} {\bibfnamefont {M.}~\bibnamefont
  {Sidler}}, \bibinfo {author} {\bibfnamefont {P.}~\bibnamefont {Back}},
  \bibinfo {author} {\bibfnamefont {O.}~\bibnamefont {Cotlet}}, \bibinfo
  {author} {\bibfnamefont {A.}~\bibnamefont {Srivastava}}, \bibinfo {author}
  {\bibfnamefont {T.}~\bibnamefont {Fink}}, \bibinfo {author} {\bibfnamefont
  {M.}~\bibnamefont {Kroner}}, \bibinfo {author} {\bibfnamefont
  {E.}~\bibnamefont {Demler}}, \ and\ \bibinfo {author} {\bibfnamefont
  {A.}~\bibnamefont {Imamoglu}},\ }\href@noop {} {\bibfield  {journal}
  {\bibinfo  {journal} {Nature Physics}\ }\textbf {\bibinfo {volume} {13}},\
  \bibinfo {pages} {255} (\bibinfo {year} {2017})}\BibitemShut {NoStop}%
\bibitem [{\citenamefont {Palzer}\ \emph {et~al.}(2009)\citenamefont {Palzer},
  \citenamefont {Zipkes}, \citenamefont {Sias},\ and\ \citenamefont
  {K\"ohl}}]{Kohl2009}%
  \BibitemOpen
  \bibfield  {author} {\bibinfo {author} {\bibfnamefont {S.}~\bibnamefont
  {Palzer}}, \bibinfo {author} {\bibfnamefont {C.}~\bibnamefont {Zipkes}},
  \bibinfo {author} {\bibfnamefont {C.}~\bibnamefont {Sias}}, \ and\ \bibinfo
  {author} {\bibfnamefont {M.}~\bibnamefont {K\"ohl}},\ }\href {\doibase
  10.1103/PhysRevLett.103.150601} {\bibfield  {journal} {\bibinfo  {journal}
  {Phys. Rev. Lett.}\ }\textbf {\bibinfo {volume} {103}},\ \bibinfo {pages}
  {150601} (\bibinfo {year} {2009})}\BibitemShut {NoStop}%
\bibitem [{\citenamefont {Catani}\ \emph {et~al.}(2012)\citenamefont {Catani},
  \citenamefont {Lamporesi}, \citenamefont {Naik}, \citenamefont {Gring},
  \citenamefont {Inguscio}, \citenamefont {Minardi}, \citenamefont {Kantian},\
  and\ \citenamefont {Giamarchi}}]{Catani2012}%
  \BibitemOpen
  \bibfield  {author} {\bibinfo {author} {\bibfnamefont {J.}~\bibnamefont
  {Catani}}, \bibinfo {author} {\bibfnamefont {G.}~\bibnamefont {Lamporesi}},
  \bibinfo {author} {\bibfnamefont {D.}~\bibnamefont {Naik}}, \bibinfo {author}
  {\bibfnamefont {M.}~\bibnamefont {Gring}}, \bibinfo {author} {\bibfnamefont
  {M.}~\bibnamefont {Inguscio}}, \bibinfo {author} {\bibfnamefont
  {F.}~\bibnamefont {Minardi}}, \bibinfo {author} {\bibfnamefont
  {A.}~\bibnamefont {Kantian}}, \ and\ \bibinfo {author} {\bibfnamefont
  {T.}~\bibnamefont {Giamarchi}},\ }\href {\doibase 10.1103/PhysRevA.85.023623}
  {\bibfield  {journal} {\bibinfo  {journal} {Phys. Rev. A}\ }\textbf {\bibinfo
  {volume} {85}},\ \bibinfo {pages} {023623} (\bibinfo {year}
  {2012})}\BibitemShut {NoStop}%
\bibitem [{\citenamefont {Chien}\ \emph {et~al.}(2015)\citenamefont {Chien},
  \citenamefont {Peotta},\ and\ \citenamefont {Di~Ventra}}]{Chien2015}%
  \BibitemOpen
  \bibfield  {author} {\bibinfo {author} {\bibfnamefont {C.-C.}\ \bibnamefont
  {Chien}}, \bibinfo {author} {\bibfnamefont {S.}~\bibnamefont {Peotta}}, \
  and\ \bibinfo {author} {\bibfnamefont {M.}~\bibnamefont {Di~Ventra}},\
  }\href@noop {} {\bibfield  {journal} {\bibinfo  {journal} {Nature Physics}\
  }\textbf {\bibinfo {volume} {11}},\ \bibinfo {pages} {998} (\bibinfo {year}
  {2015})}\BibitemShut {NoStop}%
\bibitem [{\citenamefont {Dieterle}\ \emph {et~al.}(2021)\citenamefont
  {Dieterle}, \citenamefont {Berngruber}, \citenamefont {H\"olzl},
  \citenamefont {L\"ow}, \citenamefont {Jachymski}, \citenamefont {Pfau},\ and\
  \citenamefont {Meinert}}]{Dieterle2021}%
  \BibitemOpen
  \bibfield  {author} {\bibinfo {author} {\bibfnamefont {T.}~\bibnamefont
  {Dieterle}}, \bibinfo {author} {\bibfnamefont {M.}~\bibnamefont
  {Berngruber}}, \bibinfo {author} {\bibfnamefont {C.}~\bibnamefont {H\"olzl}},
  \bibinfo {author} {\bibfnamefont {R.}~\bibnamefont {L\"ow}}, \bibinfo
  {author} {\bibfnamefont {K.}~\bibnamefont {Jachymski}}, \bibinfo {author}
  {\bibfnamefont {T.}~\bibnamefont {Pfau}}, \ and\ \bibinfo {author}
  {\bibfnamefont {F.}~\bibnamefont {Meinert}},\ }\href {\doibase
  10.1103/PhysRevLett.126.033401} {\bibfield  {journal} {\bibinfo  {journal}
  {Phys. Rev. Lett.}\ }\textbf {\bibinfo {volume} {126}},\ \bibinfo {pages}
  {033401} (\bibinfo {year} {2021})}\BibitemShut {NoStop}%
\bibitem [{\citenamefont {Gross}(1962)}]{Gross1962}%
  \BibitemOpen
  \bibfield  {author} {\bibinfo {author} {\bibfnamefont {E.}~\bibnamefont
  {Gross}},\ }\href@noop {} {\bibfield  {journal} {\bibinfo  {journal} {Annals
  of Physics}\ }\textbf {\bibinfo {volume} {19}},\ \bibinfo {pages} {234}
  (\bibinfo {year} {1962})}\BibitemShut {NoStop}%
\bibitem [{\citenamefont {Rosch}(1999)}]{rosch1999quantum}%
  \BibitemOpen
  \bibfield  {author} {\bibinfo {author} {\bibfnamefont {A.}~\bibnamefont
  {Rosch}},\ }\href@noop {} {\bibfield  {journal} {\bibinfo  {journal}
  {Advances in Physics}\ }\textbf {\bibinfo {volume} {48}},\ \bibinfo {pages}
  {295} (\bibinfo {year} {1999})}\BibitemShut {NoStop}%
\bibitem [{\citenamefont {Tomza}\ \emph {et~al.}(2019)\citenamefont {Tomza},
  \citenamefont {Jachymski}, \citenamefont {Gerritsma}, \citenamefont
  {Negretti}, \citenamefont {Calarco}, \citenamefont {Idziaszek},\ and\
  \citenamefont {Julienne}}]{Tomza-2019}%
  \BibitemOpen
  \bibfield  {author} {\bibinfo {author} {\bibfnamefont {M.}~\bibnamefont
  {Tomza}}, \bibinfo {author} {\bibfnamefont {K.}~\bibnamefont {Jachymski}},
  \bibinfo {author} {\bibfnamefont {R.}~\bibnamefont {Gerritsma}}, \bibinfo
  {author} {\bibfnamefont {A.}~\bibnamefont {Negretti}}, \bibinfo {author}
  {\bibfnamefont {T.}~\bibnamefont {Calarco}}, \bibinfo {author} {\bibfnamefont
  {Z.}~\bibnamefont {Idziaszek}}, \ and\ \bibinfo {author} {\bibfnamefont
  {P.~S.}\ \bibnamefont {Julienne}},\ }\href {\doibase
  10.1103/RevModPhys.91.035001} {\bibfield  {journal} {\bibinfo  {journal}
  {Rev. Mod. Phys.}\ }\textbf {\bibinfo {volume} {91}},\ \bibinfo {pages}
  {035001} (\bibinfo {year} {2019})}\BibitemShut {NoStop}%
\bibitem [{\citenamefont {Astrakharchik}\ \emph {et~al.}(2021)\citenamefont
  {Astrakharchik}, \citenamefont {Ardila}, \citenamefont {Schmidt},
  \citenamefont {Jachymski},\ and\ \citenamefont
  {Negretti}}]{Astrakharchik2021}%
  \BibitemOpen
  \bibfield  {author} {\bibinfo {author} {\bibfnamefont {G.~E.}\ \bibnamefont
  {Astrakharchik}}, \bibinfo {author} {\bibfnamefont {L.~A.~P.}\ \bibnamefont
  {Ardila}}, \bibinfo {author} {\bibfnamefont {R.}~\bibnamefont {Schmidt}},
  \bibinfo {author} {\bibfnamefont {K.}~\bibnamefont {Jachymski}}, \ and\
  \bibinfo {author} {\bibfnamefont {A.}~\bibnamefont {Negretti}},\ }\href@noop
  {} {\bibfield  {journal} {\bibinfo  {journal} {Communications Physics}\
  }\textbf {\bibinfo {volume} {4}},\ \bibinfo {pages} {94} (\bibinfo {year}
  {2021})}\BibitemShut {NoStop}%
\bibitem [{\citenamefont {Christensen}\ \emph {et~al.}(2022)\citenamefont
  {Christensen}, \citenamefont {Camacho-Guardian},\ and\ \citenamefont
  {Bruun}}]{Christensen-2022}%
  \BibitemOpen
  \bibfield  {author} {\bibinfo {author} {\bibfnamefont {E.~R.}\ \bibnamefont
  {Christensen}}, \bibinfo {author} {\bibfnamefont {A.}~\bibnamefont
  {Camacho-Guardian}}, \ and\ \bibinfo {author} {\bibfnamefont {G.~M.}\
  \bibnamefont {Bruun}},\ }\href {\doibase 10.1103/PhysRevA.105.023309}
  {\bibfield  {journal} {\bibinfo  {journal} {Phys. Rev. A}\ }\textbf {\bibinfo
  {volume} {105}},\ \bibinfo {pages} {023309} (\bibinfo {year}
  {2022})}\BibitemShut {NoStop}%
\bibitem [{\citenamefont {My\ifmmode~\acute{s}\else \'{s}\fi{}liwy}\ and\
  \citenamefont {Jachymski}(2024)}]{Mysliwy2024}%
  \BibitemOpen
  \bibfield  {author} {\bibinfo {author} {\bibfnamefont {K.}~\bibnamefont
  {My\ifmmode~\acute{s}\else \'{s}\fi{}liwy}}\ and\ \bibinfo {author}
  {\bibfnamefont {K.}~\bibnamefont {Jachymski}},\ }\href {\doibase
  10.1103/PhysRevB.109.214208} {\bibfield  {journal} {\bibinfo  {journal}
  {Phys. Rev. B}\ }\textbf {\bibinfo {volume} {109}},\ \bibinfo {pages}
  {214208} (\bibinfo {year} {2024})}\BibitemShut {NoStop}%
\bibitem [{\citenamefont {Pessoa}\ \emph {et~al.}(2024)\citenamefont {Pessoa},
  \citenamefont {Vitiello},\ and\ \citenamefont {Ardila}}]{Luis2024}%
  \BibitemOpen
  \bibfield  {author} {\bibinfo {author} {\bibfnamefont {R.}~\bibnamefont
  {Pessoa}}, \bibinfo {author} {\bibfnamefont {S.~A.}\ \bibnamefont
  {Vitiello}}, \ and\ \bibinfo {author} {\bibfnamefont {L.~A. P.~n.}\
  \bibnamefont {Ardila}},\ }\href {\doibase 10.1103/PhysRevLett.133.233002}
  {\bibfield  {journal} {\bibinfo  {journal} {Phys. Rev. Lett.}\ }\textbf
  {\bibinfo {volume} {133}},\ \bibinfo {pages} {233002} (\bibinfo {year}
  {2024})}\BibitemShut {NoStop}%
\bibitem [{\citenamefont {Jeblick}\ \emph {et~al.}(2017)\citenamefont
  {Jeblick}, \citenamefont {Mitrouskas}, \citenamefont {Petrat},\ and\
  \citenamefont {Pickl}}]{Jeblick2017}%
  \BibitemOpen
  \bibfield  {author} {\bibinfo {author} {\bibfnamefont {M.}~\bibnamefont
  {Jeblick}}, \bibinfo {author} {\bibfnamefont {D.}~\bibnamefont {Mitrouskas}},
  \bibinfo {author} {\bibfnamefont {S.}~\bibnamefont {Petrat}}, \ and\ \bibinfo
  {author} {\bibfnamefont {P.}~\bibnamefont {Pickl}},\ }\href@noop {}
  {\bibfield  {journal} {\bibinfo  {journal} {Communications in Mathematical
  Physics}\ }\textbf {\bibinfo {volume} {356}},\ \bibinfo {pages} {143}
  (\bibinfo {year} {2017})}\BibitemShut {NoStop}%
\bibitem [{\citenamefont {Fermi}\ and\ \citenamefont
  {Teller}(1947)}]{FermiTeller1947}%
  \BibitemOpen
  \bibfield  {author} {\bibinfo {author} {\bibfnamefont {E.}~\bibnamefont
  {Fermi}}\ and\ \bibinfo {author} {\bibfnamefont {E.}~\bibnamefont {Teller}},\
  }\href {\doibase 10.1103/PhysRev.72.399} {\bibfield  {journal} {\bibinfo
  {journal} {Phys. Rev.}\ }\textbf {\bibinfo {volume} {72}},\ \bibinfo {pages}
  {399} (\bibinfo {year} {1947})}\BibitemShut {NoStop}%
\bibitem [{\citenamefont {Mott}(1949)}]{Mott1947}%
  \BibitemOpen
  \bibfield  {author} {\bibinfo {author} {\bibfnamefont {N.}~\bibnamefont
  {Mott}},\ }\href@noop {} {\bibfield  {journal} {\bibinfo  {journal}
  {Proceedings of the Physical Society. Section A}\ }\textbf {\bibinfo {volume}
  {62}},\ \bibinfo {pages} {136} (\bibinfo {year} {1949})}\BibitemShut
  {NoStop}%
\bibitem [{\citenamefont {Arista}\ and\ \citenamefont
  {Brandt}(1983)}]{Arista1983}%
  \BibitemOpen
  \bibfield  {author} {\bibinfo {author} {\bibfnamefont {N.}~\bibnamefont
  {Arista}}\ and\ \bibinfo {author} {\bibfnamefont {W.}~\bibnamefont
  {Brandt}},\ }\href@noop {} {\bibfield  {journal} {\bibinfo  {journal}
  {Journal of Physics C: Solid State Physics}\ }\textbf {\bibinfo {volume}
  {16}},\ \bibinfo {pages} {L1217} (\bibinfo {year} {1983})}\BibitemShut
  {NoStop}%
\bibitem [{\citenamefont {Williams}(1975)}]{Williams1975}%
  \BibitemOpen
  \bibfield  {author} {\bibinfo {author} {\bibfnamefont {M.}~\bibnamefont
  {Williams}},\ }\href@noop {} {\bibfield  {journal} {\bibinfo  {journal}
  {Journal of Physics D: Applied Physics}\ }\textbf {\bibinfo {volume} {8}},\
  \bibinfo {pages} {2138} (\bibinfo {year} {1975})}\BibitemShut {NoStop}%
\bibitem [{\citenamefont {Grochowski}\ \emph {et~al.}(2020)\citenamefont
  {Grochowski}, \citenamefont {Karpiuk}, \citenamefont {Brewczyk},\ and\
  \citenamefont {Rz\k{a}\ifmmode~\dot{z}\else
  \.{z}\fi{}ewski}}]{Grochowski2020}%
  \BibitemOpen
  \bibfield  {author} {\bibinfo {author} {\bibfnamefont {P.~T.}\ \bibnamefont
  {Grochowski}}, \bibinfo {author} {\bibfnamefont {T.}~\bibnamefont {Karpiuk}},
  \bibinfo {author} {\bibfnamefont {M.}~\bibnamefont {Brewczyk}}, \ and\
  \bibinfo {author} {\bibfnamefont {K.}~\bibnamefont
  {Rz\k{a}\ifmmode~\dot{z}\else \.{z}\fi{}ewski}},\ }\href {\doibase
  10.1103/PhysRevLett.125.103401} {\bibfield  {journal} {\bibinfo  {journal}
  {Phys. Rev. Lett.}\ }\textbf {\bibinfo {volume} {125}},\ \bibinfo {pages}
  {103401} (\bibinfo {year} {2020})}\BibitemShut {NoStop}%
\end{thebibliography}%

\end{document}